\documentclass[aps,prb,twocolumn,a4paper]{revtex4}

\usepackage{graphicx}
\begin{document}
\setlength{\textheight}{255mm}

\title{Dielectrophoresis of nanoscale double-stranded DNA and humidity effects on its electrical conductivity}

\author{S.~Tuukkanen, A.~Kuzyk, J.~J.~Toppari, V.~P.~Hyt\"onen$^*$, T.~Ihalainen$^*$, and P.~T\"orm\"a}

\affiliation{Nanoscience Center, Department of Physics and $^*$Department of Biological and Environmental Science, 
P.O.Box 35 (YN), FIN-40014 University of Jyv\"askyl\"a, Finland}

\date{\today}

\begin{abstract}
Dielectrophoresis method for trapping and attaching nanoscale double-stranded
DNA between nanoelectrodes was developed. The method gives a high
yield of trapping single or a few molecules only which enables 
transport measurements at the single molecule level. Electrical
conductivity of individual 140-nm-long DNA molecules was measured, 
showing insulating behaviour in dry conditions. In contrast,
clear enhancement of conductivity was observed in moist conditions, 
relating to the interplay between the conformation of DNA molecules 
and their conductivity.
\end{abstract}

\maketitle

Controlled manipulation of single molecules is a prerequisite for fully 
understanding their properties as well as for realizing
their potential in molecular electronics. At the present, the 
fabrication of single-molecule devices in nanoscale mostly relies on
passive, uncontrollable methods of manipulation such as deposition of 
the molecules on the substrate or on the fabricated
structure. Dielectrophoresis$^{1,2}$ (DEP), an active manipulation 
method utilizing electro-magnetic fields, has been widely applied 
for microscale objects,$^3$ e.g., DNA of bacteriophage lambda ($\lambda$-DNA).$^{4-6}$ 
In nanoscale, however, Brownian motion poses a challenge:~the few
successful demonstrations are for trapping nanoscale objects,$^{7,8}$ 
and for attaching DNA molecules between nanoelectrodes by
DC-DEP.$^9$ Concerning the intriguing question of DNA 
conductivity,$^{9-18}$ there starts to be a consensus that 
double-stranded DNA (dsDNA) molecules
exposed to untreated SiO$_2$ or mica surfaces, in dry environment or 
vacuum, are insulating.$^{19-23}$ However, the conductivity
of DNA on specially treated surfaces,$^{23,24}$ in solutions$^{25-27}$ 
or inside dried films$^{10}$ remains open. Also, the effect
of humidity on the electrical conductivity of DNA films$^{28,29}$ 
or constellations of DNA molecules$^{30,31}$ has been discussed
recently. The effects of the ambient conditions are related to the 
intimate connection between the conformation of
the molecules and their conductivity.

In the present paper, we report a fully developed AC-DEP technique applicable for trapping, 
stretching and attaching nanoscale dsDNA molecules between nanoelectrodes. 
The technique has a high yield and allows transport measurements of single or a few molecules. 
Electrical conductivity of the trapped, 140 nm long dsDNA molecules was measured. 
Especially, the effect of humidity was investigated. 
While dsDNA in dry environment showed insulating behaviour, the molecules in moist conditions showed
significantly lower resistances (linear resistance of the order of 100 M$\Omega$) providing the first observation of 
humidity effects for individual nanoscale DNA molecules.

%The dielectrophoretic force decreases with the length of the molecule and, 
%in nanoscale, high gradients of the electric field are needed.$^{1}$ 
We fabricated narrow finger-tip type gold electrodes, with a gap of about 
100 nm, on a SiO$_2$ substrate using standard electron beam 
lithography (Fig.~1; see EPAPS Ref.~32).
We chose to use AC-DEP instead of DC to eliminate undesired electrophoretic effects
and to enhance stretching of dsDNA molecules.$^4$ 
Double-stranded 414 bp ($\sim$140 nm) long DNA containing a 
thiol group (--SH) in both ends was fabricated and 
diluted in Hepes buffer.
%$^{33}$
%DEP experiments were performed by incubating a few microliter drop 
%of the DNA solution onto the sample and applying a sinusoidal voltage to electrodes.$^{32}$
To optimize the process, we studied the DEP of fluorescent
labeled DNA {\it in situ} under a confocal microscope
%, with frequencies $0.1-20$ MHz and AC-voltages $1-4$ V (peak-to-peak).
%The higher the frequency was the more efficiently the DNA was localized to the area of the highest 
%field gradient, i.e., near the gap 
[Fig.~1(a); see EPAPS Ref.~32 and the movies in Ref.~33].
%The use of high frequencies was limited by the higher voltages needed, and
The optimal DEP frequency was found to be 750 kHz  
combined with field strength of $\sim$10$^7$ V/m.
%In Fig.~1b--d are shown atomic force microscope (AFM) images of samples with one, 
%two and three DNA helices trapped between the electrodes.

Electrical DC conductivity measurements of the DNA were done in 
room temperature (23$^\circ$C) both with relative air humidity of
about 30 \% ('dry' environment) and of 80\%-90\% ('moist' environment).
Tens of samples containing DNA were measured in the dry environment,
and they all showed insulating behaviour: $I$-$V$ curves were  
linear at small voltages with resistance of about 10 T$\Omega$.\linebreak
\begin{widetext}
\begin{figure}[bh]
\parbox{\textwidth}{
\includegraphics[width=\textwidth]{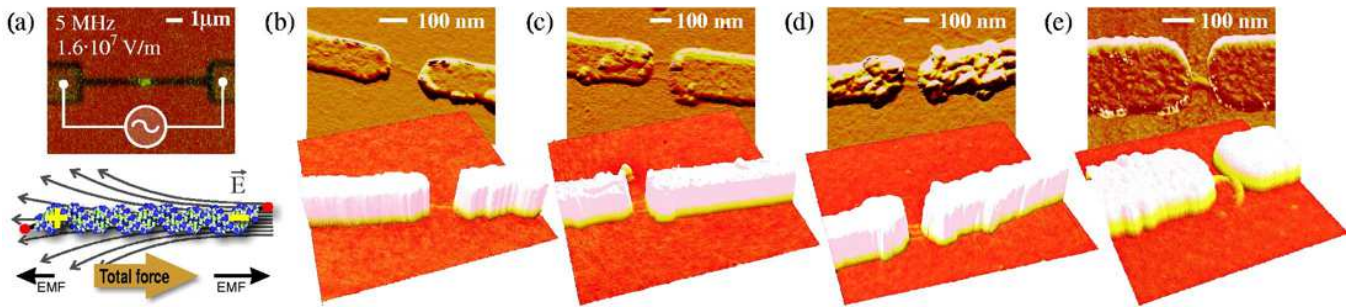}
\caption{(a) Dielectrophoresis under confocal microscope (after 1 min),
and the principle of DEP. AFM pictures of one (b), two 
(c), and three (d) (Sample I) DNA molecules, and a DNA bundle (e) (Sample II) trapped 
between electrodes using DEP. The heights of the 
individual molecules were $\sim$1 nm when measured in dry 
environment and the bundle was $\sim$6 nm
high containing thus only a few molecules (See EPAPS Ref.~32).   
%Some of the electrodes are cut at the top in the images.
} \label{fig1}}
\end{figure}
\end{widetext}
These resistance values for dry dsDNA on the SiO$_2$ surface are in agreement with many 
recent observations by other groups.$^{10,11,14,19-22,24,31}$ 
In contrast, in moist conditions, several samples showed clear increase in 
conductivity which was much higher than observed in the reference 
samples, for which control experiments were done using exactly the same procedure 
for the DEP and subsequent transport measurements, but using a 
buffer solution without DNA.

For instance, a sample showing conductivity in humid air was the 
one with three individual DNA molecules, Fig.~1(d) (Sample I). In
dry environment, the resistance was $\sim$10 T$\Omega$. It dropped 
to $\sim$250 M$\Omega$ after the sample had been half an hour
in moist environment [red circles in Fig.~2(a)]. After that, the 
resistance slowly increased during the measurement, resulting to
$\sim$700 M$\Omega$ after three hours (blue open circles). This 
deterioration of conductivity during the measurements is probably
due to disturbance of the DNA structure caused by gathering of 
contaminants from the moist air.$^{32}$
After the measurement in moisture, the sample was dried with 
nitrogen and the resistance increased back to the original dry value.
The sample was imaged with AFM right after these measurements 
confirming that at least two of the DNAs were still properly
attached [inset in Fig.~2(a)]. The material between the electrodes 
was confirmed afterwards to be dsDNA using confocal microscopy with
dsDNA-specific fluorescent labelling.$^{32}$ 
Similar behaviour was observed also in a sample containing a bundle of DNA, Fig.~1(e)
(Sample II). The resistance was a few T$\Omega$ in dry 
environment and $\sim$40 G$\Omega$ immediately after applying the
moist conditions, but decreased to $\sim$250 M$\Omega$ after the sample 
had been over ten hours at moist conditions [circles in Fig.~2(b)]. The 
increase in conductivity in this case was much slower than in the case 
of Sample I, furthermore, the resistance stayed the same during 
the measurements and did not increase as in the case of Sample I. The
sample was finally dried and the resistance rose to a few T$\Omega$ 
again. The AFM image in inset of Fig.~2(b) shows that the DNA
bundle was still in place after the measurement. 

The behaviour of Samples I and II, i.e., resistance dropping 
to hundreds of M$\Omega$ in moist conditions, was observed in five 
different samples with single or a few DNAs. Such behaviour was 
never observed in the reference samples, containing no DNA. 
However, some of the samples containing DNA behaved in a similar way to 
reference samples, indicating that there is either no 
DNA properly attached to the electrodes or the DNA is not 
conducting, e.g., due to being severely deformed.  
$I$-$V$ curves from one of the reference samples are
shown in Fig.~2(c). They also show clear difference between the 
dry and moist environment measurements. However, in moist
conditions, the minimum resistance observed for the reference 
samples was $\sim$7 G$\Omega$, and the resistance in dry
environment was always around 10 T$\Omega$. The number 
of samples, the double check with AFM and confocal imaging, and the
comparison to the reference samples using buffer without DNA 
provide, altogether, firm evidence for the strong effect of moisture
on the electrical conductivity of single nanoscale DNA molecules.
Note that the conductance can still be limited by the used {\it hexanethiol}\/-linkers 
reported resistance of $10^7 - 10^9$ $\Omega$.$^{34}$

\begin{figure}[t]
\includegraphics[width=55 truemm]{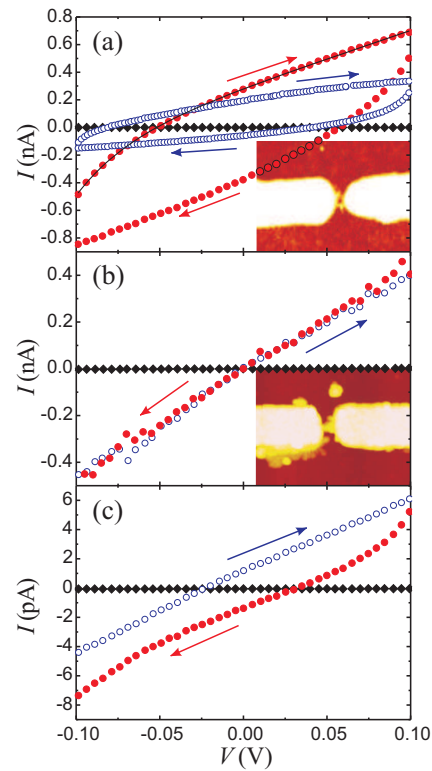}
\caption{$I$-$V$ characteristics of Sample I [Fig.~1(d)] in (a), of 
Sample II [Fig.~1(e)] in (b) and, in (c), of one of the
reference samples without DNA. In all, the black diamonds are recorded 
in dry environment and the circles are from measurements at
moist conditions. The insets show AFM images of the 
samples taken right after the measurements. The hysteretic behaviour is 
due to enhanced charging effects in moisture, 
as shown by fitting to theory including such effects (solid curve).}
\label{fig2}
\end{figure}

Even when the effect of humidity on the conductivity of 
individual DNA molecules is evident, the nature of the charge transport
cannot be completely determined based on these experiments. In 
earlier experiments,$^{28-31}$ the humidity enhanced conductivity
of DNA has been explained by dipole relaxation losses of,$^{29}$ or 
dissociation, i.e., proton transfer through,$^{28,30}$ the hydrated 
water molecules. The first model applies only to AC conductivity. 
Since reduction-oxidation processes are negligible due to 
low voltages used in our experiments,$^{27}$ buffer salts 
and the counterions do not contribute to the total steady-state DC-current.
Instead, diffusion of the ions to the electrodes, especially in moist 
environment, causes extra capacitance 
%. This increases charging effects 
%during the measurements and induces hysteresis 
as seen in Fig.~2.
One more possibility is enhanced electron transport/transfer caused
by humidity induced conformational changes in DNA structure. 
The direct electronic conductivity, by means of overlapping
$\pi$-orbitals of the base pairs along the molecular axis, is 
likely to be sensitive to the helical conformation of dsDNA (Refs. 35 and 36) 
%$^{39,40}$)
(the contributions of protons or counterions might also be affected by 
the deformations). Also magnetic properties of 
$\lambda$-DNA are shown to depend on the conformation of dsDNA.$^{37}$
The deformations of the structure can be due to, e.g., ambient conditions 
such as humidity, or interactions with the substrate surface.$^{23}$ 
For instance, a single dsDNA on graphite appears in its natural B-DNA form at 
moist conditions, but collapses to a form resembling A-DNA (defected 
overlap of the $\pi$-orbitals) when dried to the surface.$^{38}$
[We observed, in dry conditions, reduced height of the dsDNA
($\sim$1 nm compared to the expected 2 nm) corresponding to a deformed state.]
Also the contribution of the positive counterions to
the electrical conductivity via gating effect has been suggested.$^{39}$

In our case, the slower time scale of the conductivity change for 
DNA bundle (Sample II) vs.~single molecules suggests that
proton transfer along the dissociated hydrated water layer is not 
dominant: it would be enough to have water present at the
surface of the object, which should happen as fast for a bundle 
as for a single molecule. The slower conductivity change in 
Sample II could be a result of slower hydrations of DNA 
helices inside the bundle. Likewise, the bundle
can also keep the moisture inside and protect inner molecules, 
thus making the B-state more stable, as observed in the
measurements. In the samples with individual molecules, 
the assumed recovery of the helical conformation due to humidity was
either much faster, e.g., Sample I, or not successful at all. 
Also the increase of conductance at moist conditions only in some
of the samples suggests the charge transfer mechanism being 
highly sensitive to the conformation, rather than proton transfer
through the water layer on a (deformed) molecule. Our 
results are consistent with the resistance values observed
for $\lambda$-DNA (Ref. 25) and short duplexes$^{27}$ in buffer. This 
suggests that the hydration layer around the dsDNA in high
humidity environment enables similar behaviour than the buffer 
environment, e.g., maintaining the double helical conformation of
B-DNA. 

In summary, we have developed a nanoscale AC-DEP technique that 
has a high yield and provides a platform for reliable transport
measurements at the single molecule level. We observed a remarkable 
increase in the electrical conductivity of 140 nm long (414
bp) dsDNA molecules with increasing humidity of ambient air, and 
the observation was confirmed by various reliability
checks.$^{32}$ Our results also suggest that the change is related 
to a humidity induced conformational change of the molecular
structure and associated with a contribution from electron transfer. 
Further research is required, however, to identify in detail
the contributions from electrons, water ions and counterions.

{\it Acknowledgements}. 
The authors thank J.~A.~Virtanen and M.~S.~Kulomaa 
for useful discussions and acknowledge the financial support
by Academy of Finland (Project Nos.~205470, 53903), the Emil Aaltonen 
foundation and the National Graduate School in Informational and
Structural Biology (V.P.H.).

%{\bf EPAPS Available:} Descriptions of experimental 
%procedures and setups used.


\begin{thebibliography}{99}
\footnotesize

\bibitem{1} H.~A.~Pohl, J.~Appl.~Phys. {\bf 22}, 869 (1951); 
H.~A.~Pohl, {\it Dielectrophoresis: The Behavior of
Neutral Matter in Nonuniform Electric Fields} (Cambridge 
University Press, Cambridge, UK, 1978).

\bibitem{burkereview} For a review, see e.g., P.~J.~Burke, 
Encyclopedia of Nanoscience and Nanotechnology, Vol.~6, p.~623 (2004).

\bibitem{microdep}P.~Debye, P.~P.~Debye, and B.~H.~Eckstein, Phys.~Rev. {\bf 94}, 1412 (1954); 
J.~S.~Batchelder, Rev.~Sci.~Instrum. {\bf 54}, 300 (1983); 
F.~F.~Becker, X.-B.~Wang, Y.~Huang, R.~Pethig, J.~Vykoukal, P.~R.~C.~J.~Gascoyne, 
J.~Phys.~D {\bf 27}, 2659 (1994).

\bibitem{washizu} M.~Washizu and O.~Kurosawa, IEEE 
Trans.~Indust.~Appl. {\bf 26}, 1165 (1990); 
%M.~Washizu, O.~Kurosawa, I.~Arai, S.~Suzuki, N.~Shimamoto, IEEE Trans.~Indust.~Appl. {\bf 31}, 447 (1995);
S.~Suzuki, T.~Yamanashi, S.~Tazawa, O.~Kurosawa, M.~Washizu,
{\it ibid.} %IEEE Trans.~Indust.~Appl. 
\textbf{34}, 75 (1998).

\bibitem{5} C.~L.~Asbury and G.~van den Engh, Biophys.~J. 
{\bf 74}, 1024 (1998); 
S.~Tsukahara, K.~Yamanaka, and H.~Watarai, Chem.~Lett. {\bf 3}, 250 (2001);
W.~A.~Germishuizen, C.~W\"alti, R.~Wirtz, M.~B.~Johnston, M.~Pepper, 
A.~G.~Davies, A.~P.~J.~Middelberg, Nanotechnology {\bf 14}, 896 (2003); 
L.~Zheng, J.~P.~Brody, and P.~J.~Burke, Biosensors Bioelectron. {\bf 20}, 606 (2004).

\bibitem{hartzell} B.~Hartzell,
B.~McCord, D.~Asare, H.~Chen, J.~J.~Heremans,  V.~Soghomonian,
Appl.~Phys.~Lett. {\bf 82}, 4800 (2003); 
%B.~Hartzell, B.~McCord, D.~Asare, H.~Chen,J.~J.~Heremans, V.~Soghomonian,
%J.~Appl.~Phys. {\bf 94}, 2764 (2003).

\bibitem{nanodep} %DEP applied in nanoscale for: Proteins: 
M.~Washizu, S.~Suzuki, O.~Kurosawa, T.~Nishizaka, T.~Shinohara,
IEEE Trans.~Indust.~Appl. {\bf 30}(4), 835 (1994); 
%Viruses: 
T.~Schnelle, S.~Muller, S.~Fiedler, G.~Shirely, K.~Ludwig, A.~Hermann, G.~Fuhr,
Naturwissenschaften, {\bf 83}, 172 (1996); 
%Nanoparticles: 
A.~Bezryadin, C.~Dekker, and G.~Schmid,
Appl.~Phys.~Lett. {\bf 71}, 1273 (1997); 
%Latex spheres: 
N.~G.~Green, H.~Morgan, and J.~J.~J.~Milner,
Biochem.~Biophys.~Methods {\bf 35}, 89 (1997); 
%Carbon nanotubes: 
X.~Q.~Chen, T.~Saito, H.~Yamada, K.~Matsushige,
Appl.~Phys.~Lett. {\bf 78}, 3714 (2001).

\bibitem{nanodnadep} %DEP applied in nanoscale for DNA, no attachment to electrodes: Electrodeless DEP: 
C.-F.~Chou, J.~Tegenfeldt, O.~Bakajin, S.~S.~Chan,
E.~C.~Cox, N.~Darnton, T.~Duke, and R.~H.~Austin,
Biophys.~J. \textbf{83}, 2170 (2002); 
%DEP in a nanopipette: 
L.~Ying, S.~S.~White, A.~Bruckbauer, L.~Meadows,
Y.~E.~Korchev, and D.~Klenerman,
Biophys.~J. \textbf{86}, 1018 (2004).

\bibitem{porath}D.~Porath,
A.~Bezryadin, S.~De Vries, and C.~Dekker,
Nature \textbf{403}, 635 (2000).

\bibitem{okahata} Y.~Okahata,
T.~Kobayashi, K.~Tanaka, and M.~Shimomura,
J.~Am.~Chem.~Soc. \textbf{120}, 6165 (1998); Y.~Okahata,
T.~Kobayashi, H.~Nakayama, and K.~Tanaka,
Supramol.~Sci. \textbf{5}, 317 (1998).

\bibitem{braun} E.~Braun,
Y.~Eichen, U.~Sivan, and G.~Ben-Yoseph,
Nature \textbf{391}, 775 (1998).

\bibitem{kelley} S.~O.~Kelley and J.~K.~Barton, Science \textbf{283}, 375 (1999).

\bibitem{fink} H.-W.~Fink and C.~Sch\"onenberger, 
Nature \textbf{398}, 407 (1999).

\bibitem{pablo} P.~J.~de~Pablo,
F.~Moreno-Herrero, J.~Colchero, J.~Gómez-Herrero,
P.~Herrero, A.~M.~Baro, P.~Ordejon, J.~M.~Soler, and E.~Artacho,
Phys.~Rev.~Lett. \textbf{85}, 4992 (2000).

\bibitem{cai}L.~Cai, H.~Tabata, and T.~Kawai, 
Appl.~Phys.~Lett. \textbf{77}, 3105 (2000).

\bibitem{kasumovscience}A.~Yu.~Kasumov,
M.~Kociak, S.~Guéron, B.~Reulet,
V.~T.~Volkov, D.~V.~Klinov, and H.~Bouchiat,
Science \textbf{291}, 280 (2001).

\bibitem{yoo}K.-H.~Yoo,
D.~H.~Ha, J.-O.~Lee, J.~W.~Park, J.~Kim,
J.~J.~Kim, H.-Y.~Lee, T.~Kawai, and H.~Y.~Choi,
Phys.~Rev.~Lett. \textbf{87}, 198102 (2001).

\bibitem{rakitin} A.~Rakitin,
P.~Aich, C.~Papadopoulos, Yu.~Kobzar,
A.~S.~Vedeneev, J.~S.~Lee, and J.~M.~Xu,
Phys.~Rev.~Lett. \textbf{86}, 3670 (2001).

\bibitem{storm} A.~J.~Storm,
J.~van Noort, S.~de Vries, and C.~Dekker,
Appl.~Phys.~Lett.  \textbf{79}, 3881 (2001).

\bibitem{zhang} Y.~Zhang,
R.~H.~Austin, J.~Kraeft, E.~C.~Cox, and N.~P.~Ong,
Phys.~Rev.~Lett. \textbf{89}, 198102 (2002).

\bibitem{bockrath} M.~Bockrath,
N.~Markovic, A.~Shepard, m.~Tinkham, L.~Gurevich,
L.~P.~Kouwenhoven, M.~W.~Wu, and L.~L.~Sohn,
Nano Lett. \textbf{2}, 187 (2002).

\bibitem{gomez}C.~G\'omez-Navarro,
F.~Moreno-Herrero, P.~J.~de Pablo, J.~Colchero, J.~G\'omez-Herrero, and A.~M.~Bar\'o,
Proc.~Natl.~Acad.~Sci.~U.S.A. \textbf{99}, 8484 (2002).

\bibitem{kasumovAPL}A.~Yu.~Kasumov,
D.~V.~Klinov, P.-E.~Roche, S.~Gu\'eron, and H.~Bouchiat,
Appl.~Phys.~Lett. \textbf{84}, 1007 (2004).

\bibitem{heim}T.~Heim, D.~Deresmes, and D.~Vuillaume,
J.~Appl.~Phys. \textbf{96}, 2927 (2004).

\bibitem{tran}P.~Tran, B.~Alavi, and G.~Gr\"uner,
Phys.~Rev.~Lett. \textbf{85}, 1564 (2000).

\bibitem{boon} E.~M.~Boon and J.~K.~Barton,
Curr.~Opin.~Struct.~Biol. \textbf{12}, 320 (2002).

\bibitem{xu} B.~Xu,
P.~Zhang, X.~Li, and N.~Tao,
Nano Lett. \textbf{4}, 1105 (2004).

\bibitem{otsuka} Y.~Otsuka,
H.-Y.~Lee, J.-H.~Gu, J.-O.~Lee, K.-H.~Yoo,
H.~Tanaka, H.~Tabata, and T.~Kawai,
Jpn.~J.~Appl.~Phys. \textbf{41}, 891 (2002).

\bibitem{briman} M.~Briman,
N.~P.~Armitage, E.~Helgren, and G.~Gr\"uner,
Nano Lett. \textbf{4}, 733 (2004).

\bibitem{ha}D.~H.~Ha,
H.~Nham, K.-H.~Yoo, H.-M.~So, H.-Y.~Lee, and T.~Kawai,
Chem.~Phys.~Lett. \textbf{355}, 405 (2002).

\bibitem{kleine}H.~Kleine,
R.~Wilke, Ch.~Pelargus, K.~Rott,
A.~P\"uhler, G.~Reiss, R.~Ros, and D.~Anselmetti,
J.~Biotechnol. \textbf{112}, 91 (2004).

\bibitem{support} See EPAPS Document No. E-APPLAB-87-010543 for details of the experiments. 
This document can be reached via a direct link in the online article's HTML 
reference section or via the EPAPS homepage (http://www.aip.org/pubservs/epaps.html)

%\bibitem{added}
%5'-thiol-modified oligonucleotides$^{32,34}$ were used as primers in the PCR using chicken avidin complementary 
%DNA$^{35}$ as a
%template. The final concentration was 10 nM (2.7 $\mu$g/ml) in 6.5 mM Hepes buffer (pH 5.45). 2 mM NaBH$_4$ was added 
%to cleave S-S bonds between DNA (final pH 7.0).

%\bibitem{tuukkanen} S.~Tuukkanen,
%J.~Virtanen, V.~P.~Hyt\"onen, M.~S.~Kulomaa, and P.~T\"orm\"a,
%Rev.~Adv.~Mater.~Sci. \textbf{5}, 228 (2003).

%\bibitem{gope} M.~L.~Gope,
%R.~A.~Kein\"anen, P.~A.~Kristo, O.~M.~Conneely,
%W.~G.~Beattie, T.~Zarucki-Schulz, B.~W.~O'Malley, and M.~S.~Kulomaa,
%Nucleic Acids Res. \textbf{15}, 3595 (1987).

\bibitem{www} 
http://www.phys.jyu.fi/research/electronics/research/ \\ 
depmovies.html

\bibitem{xu2}B.~Xu and N.~J.~Tao, 
Science {\bf 301}, 1221 (2003) and references therein.

\bibitem{eley}D.~D.~Eley and D.~I.~Spivey, 
Trans.~Faraday Soc. \textbf{58}, 411 (1962).

\bibitem{endress} For a review, see e.g., R.~G.~Endres, D.~L.~Cox,and R.~R.~P.~Singh,
Rev.~Mod.~Phys.~\textbf{76}, 195 (2004).

\bibitem{nakamae} S.~Nakamae,
M.~Cazayous, A.~Sacuto, P.~Monod, and H.~Bouchiat,
Phys.~Rev.~Lett.~{\bf 94}, 248102 (2005).

\bibitem{zareie} M.~H.~Zareie and P.~B.~Lukins,  
Biochem.~Biophys.~Res.~Commun. \textbf{303}, 153 (2003).

\bibitem{barnett}R.~N.~Barnett,
C.~L.~Cleveland, A.~Joy, U.~Landman, and G.~B.~Schuster,
Science \textbf{294}, 567 (2001).


\end{thebibliography}
\end{document}